# Experimental probing of the anisotropy of the empty p states near the Fermi level in MgB$_2$

R. F. Klie, Y. Zhu

*Brookhaven National Laboratory, Upton, New York 11973*

G. Schneider

*Institut fuer Allgemeine Physik und Center for Computational Materials Science (CMS)*

*Technische Universitaet Wien, A - 1060 WIEN, Austria*

J. Tafto

*Department of Physics, University of Oslo, P.O.Box 1048 Blindern, 0316 Oslo, Norway*

**Abstract**

We have studied the Boron K-edge in the superconductor MgB$_2$ by electron energy loss spectroscopy (EELS) and experimentally resolved the empty p states at the Fermi level that have previously been observed within an energy window of 0.8eV by soft x-ray absorption spectroscopy. Using angular resolved EELS, we find that these states at the immediate edge onset have p$_{xy}$ character in agreement with predictions from first-principle electronic structure calculations.

Since the discovery of the unexpectedly high superconducting transition temperature of 39 K[1], the compound $MgB_2$ has received much attention. According to ground state electronic structure calculations, the states near the Fermi level are the states at the boron atoms with predominantly p character.[2-4] The density functional theory (DFT) calculations (fig. 1)[4] suggest that the $p_{xy}$ states, in this highly anisotropic compound, have a high density up to 0.8 eV above the Fermi level before the density of states drops to near zero and then starts to rise again about 5 eV above the Fermi level. These incompletely filled $p_{xy}$ states at the Fermi level onset are believed to play an important role in the superconductivity of this highly two-dimensional electron-hole BCS superconductor. Including the core-hole effect in the DFT calculation (fig. 1) will primarily lead to a decrease of the high density of $p_{xy}$ states at the Fermi level onset. The states with $p_z$-character exhibit a flat density of states over the range of 10 eV from the Fermi level onset. According to the dipole selection rule, these p states at the K edge of boron are being probed using x-ray absorption spectroscopy (XAS) and electron energy loss spectroscopy (EELS). Angle averaged soft x-ray absorption spectroscopy of the K-edge of boron exhibits, at the Fermi level, a peak consistent with a high density of p states.[3,4] By using polarized x-rays one can determine the symmetry of these states i.e. whether they are $p_{xy}$ or $p_z$ states. However, these experiments require large single crystals and to date no x-ray absorption experiments have been performed on these crystals.

Angle-resolved EELS, in which high energy electrons that have undergone momentum transfer by propagating through a thin specimen of thickness less than 100 nm are analyzed, is an alternative to x-ray absorption spectroscopy. Although in comparison to x-ray absorption spectroscopy, conventional transmission electron microscopes (TEM) lack energy resolution, the increased spatial resolution of EELS provides an invaluable tool for probing the local electronic structure of poly-crystalline materials. In their study of micrometer sized thin areas of graphite and boron nitride using a transmission electron microscope equipped with a Wien filter, Leapman et al.[5] achieved high angular resolution. With a dedicated instrument, they simultaneously observed the EELS spectrum as a function of angle and energy loss. However, large areas of sample material with comparable sample thickness and crystal orientation are necessary to perform this kind of experiments, and $MgB_2$ does not have prominent cleavage planes. Hence, relying on bulk specimens that have been thinned by ion milling or other means, it is difficult to find areas larger than 100nm in diameter that are thin enough for EELS studies. Thus, in order to obtain a good signal to noise ratio we must to sacrifice the very high angular resolution.

Several recent EELS studies[4,6-8] have aimed at obtaining angle resolved spectra of $MgB_2$ for the purpose of separating the $p_{xy}$ and the $p_z$ states by energy analyzing the electrons that continue in the forward direction after they have transmitted crystal regions less than 50 nm . For this compound with hexagonal symmetry (space group: P6/mmm) the focus has been on two incident beam directions, the incident beam parallel and perpendicular to the c-axis of. It is crucial under these conditions that the angular spread of the incident beam as well as the collected electrons are small. To achieve a high orientation sensitivity and to minimize the momentum transfer perpendicular to the electron beam direction, collections angles of less than 1 mrad for 100keV electrons at the Boron K-edge of energy about 200eV are required, and these angles need to be even smaller at higher acceleration voltages. Because of these demands there is so far no unambiguous experimental determination of the symmetry of the p states just above the Fermi level in $MgB_2$.

The EELS results in this report were obtained using the JEOL-3000F field emission STEM/TEM, equipped with a Schottky field-emission source operated at 300keV, an ultra high resolution objective lens pole piece, and a post column Gatan imaging filter (GIF). The microscope and spectrometer were setup for TEM with a "parallel illumination", where the convergence angle ($\alpha$) is chosen to be less than 0.4mrad and the spectrometer collection angle $\theta_c$ variable between 0.4 and 1.5 mrad. The EEL spectra of the B K-edge shown in this report are the sum of up to 15 individual spectra, added to increase the signal-to-noise ratio of the near-edge fine-structure. Each spectrum was acquired with an acquisition time of 2s to 5s per spectrum, a dispersion of 0.2 eV/channel and subsequently background subtracted. The experimental energy resolution of the spectra was 1.2 eV in the vacuum.

Fig.2 shows the K-edge of boron with the incident beam parallel to the c-axis for three different effective collection angles of 0.4, 0.7 and 1.5 mrad centered at the forward direction. Fig. 3 shows similar spectra with the incident beam perpendicular to the c-axis. We note that the shoulder at the edge onset decreases with increasing collection angle in fig. 2, with the incident beam along the c-axis, whereas the trend is opposite in fig. 3 with the incident beam normal to the c-axis. Furthermore, in fig. 2 the peak located at 203 eV in the 1.5 mrad spectrum shifts towards the lower end of the spectrum with decreasing collection angle. The opposite can be observed in fig. 3, where the peak shifts from 201 eV in the 1.5 mrad spectrum to 202.5 eV in the 0.4 mrad spectrum. All these features can be understood by considering the momentum transfer selection of the spectrometer entrance aperture, shown in fig 4 and the DFT calculations of the Boron K-edge (fig. 1)[4]. Figure 1 shows that the $p_z$ states have a nearly uniform and rather high density of states at the Fermi level, extending much more than 5 eV, while the $p_{xy}$ states within this energy range have the narrow high peak confined to an energy window of 0.8 eV at the Fermi level. The highest density of $p_z$ states can be found at 201 eV, whereas the highest density of $p_{xz}$-states is located at 203 eV.

Fig. 4 shows the contribution of $p_z$ and $p_{xy}$ by the momentum transfer parallel and perpendicular to the incident beam direction as a function of scattering angle. Note that these calculations are for discrete scattering directions; to calculate the spectrum for a finite collection angle we have to consider the cross section variation with scattering angle and integrate over the range of scattering angles.[8] With the incident beam along the c axis, as is the case in fig. 2, we thus predominantly probe the $p_z$ states for collection angles smaller than 0.5 mrad, while the contribution from $p_{xy}$ increases with collection angle. The trend is opposite in fig. 3 with the incident beam normal to the c-axis. Note, however, the in-plane anisotropy with the incident beam normal to the c-axis resulting for large scattering angles that the contribution from $p_z$ and $p_{xy}$ both approach 50% in fig. 4, assuming we average over the azimuthal angle as we need to do if the incident beam is in the center of the entrance aperture. On the other hand, for incident beam direction parallel to the c axis, the $p_{xy}$ contribution approach 100% at large scattering angle. Even at 1 mrad, an angle well below the regime where the dipole approximation starts to break down, the $p_{xy}$ contribution is already 90%.

Therefore, we may achieve much larger orientation selectivity by displacing the incident beam so that we collect electrons that are scattered at least 0.5 mrad. The entrance aperture of the spectrometer can now be chosen to correspond to a collection angle in excess of 0.4 mrad. Figure 5 shows two spectra that fulfill these experimental conditions. Spectrum a) is taken with the incident beam perpendicular to the c-axis, and the center of the entrance aperture is displaced by $\theta_D=1.2$ mrad towards the (001) diffraction sort; the collection angle ranges $\theta_c=0.7$mrad to $\theta_c=1.7$mrad. Due to the anisotropy of $MgB_2$ and the scattering momentum transfer, EELS for this crystal orientation will contain contribution from both the $p_{xy}$ and the $p_z$ components. By displacing the aperture by 1.2 mrad toward the 001 diffraction spot, a majority of the spectral weight from the $p_z$ contributions is chosen. A pure spectrum of the $p_z$-states in $MgB_2$ is not possible by off-axis EELS due to the required minimum aperture size. The shape of the B K-edge exhibits a high intensity pre-peak that remains constant over the range of 5 eV, which is in agreement with the first-principle calculations. Although the DFT calculation (fig. 1) suggest a featureless spectrum for energies up to 15 eV above the Fermi-level, two distinct peaks can be seen in the experimental spectrum. The intensity at 194 eV stems from the remaining $p_{xy}$ contribution in this spectrum, whereas the peak at 201 eV corresponds to the $p_z$-peak seen in the DFT calculations (fig. 1).[4]

Spectrum b) shows the results of the displaced entrance aperture with the c-axis parallel to the incoming electron beam. The displacement of the aperture is chosen to be $\theta_D=1.2$ mrad and the collection angular range of $\theta_c$=0.7-1.7mrad was selected. With the displaced aperture more than 90 % of the total spectral contribution stems from the transitions into the $p_{xy}$ states. We clearly see the pre-edge peak with a high intensity at the edge-onset and a subsequent drop in intensity, as predicted from the first-principle calculations. The width of the pre-peak of 1.4 eV, can be easily explained as the convolution of the intrinsic peak-width of 0.8 eV and the energy-spread of the incoming electron beam. The first peak above the pre-peak s located at 194 eV, and the highest peak is positioned at 203 eV. A comparison with the DFT calculations (fig. 1) shows that the high intensity at the edge onset and the relative intensities of the other peaks in the B K-edge can only be explained by not taking the core-hole effect into consideration. Including the core-hole will lead to a suppression of the initial density of states at the Fermi-level and to areceral of the peak intensities located at 194 eV and 203 eV.

In summary, we demonstrated in this paper that the pre-peak of the Boron K-edge in $MgB_2$, which was previously mainly attributed to the $p_z$ states, contains for most crystal orientation and acquisition conditions contributions from both the $p_{xy}$ and the $p_z$ components. By choosing the appropriate acquisition conditions, i.e. small collection angle and off-axis spectroscopy, the individual components of the Boron K-edge pre-peak can be directly separated. The shoulder at the Boron K-edge threshold, which is attributed to the $p_z$-states decreases with the momentum-transfer perpendicular to the c-axis, whereas a sharp peak just at the Fermi level increases with increasing fraction of the momentum transfer normal to the c axis. This leaves little doubt that the peak, previously observed by XAS at the Fermi level [4] has $p_{xy}$ character in agreement with first-principle calculations without a core-hole effect.

Concluding, we have shown in this paper that the individual contributions of the Boron K-edge pre-peak can be directly separated by EELS. We have further shown that the $p_{xy}$-peak at the Fermi level can be resolved in the experimental spectra, even with an energy resolution of more than 1.0 eV. This will allow us to compare the resulting EEL spectra directly with the different kind of DOS-calculations, and evaluate the importance of the transition matrix elements for a match between experiments and calculations. Further, the effects of dopants, point defects and grain boundaries on the local superconducting charge carrier concentration can now be evaluated directly by EELS on the nanometer scale.

This work is supported by the U.S. Department of Energy, Division of Materials Sciences, Office of Basic Energy Science, under Contract No. DE-AC02-98CH10886.

Figure captions

Fig. 1. DFT ground state calculations of the density of boron $p_{xy}$ and $p_z$ above the Fermi level (bottom). The core-hole effect is included in the top calculations.

Fig. 2. EELS spectra recorded with the incident beam along the c axis entering in the center of the entrance aperture of the spectrometer with effective collection angles of 0.4, 0.7 and 1.5 mrad.

Fig. 3. EELS spectra recorded with the incident beam perpendicular to the c axis entering in the center of the entrance aperture of the spectrometer with effective collection angles of 0.4, 0.7 and 1.5 mrad.

Fig. 4. Fractional contribution from pz and pxy as a function of scattering angle with the incident beam k parallel to the c axis (k||c) and incident beam perpendicular to the c axis (k_|_ c). With the incident beam perpendicular to the c axis these curves depend on azimutal angle. In this figure we have averaged over the azimutal angle.

Fig. 5. Spectrum with the incident beam along the c axis. The diameter of the entrance aperture corresponds to an angle of 1.5 mrad and is displaced from the forward direction so that the minimum scattering angle is 1 mrad.

Figure 1:

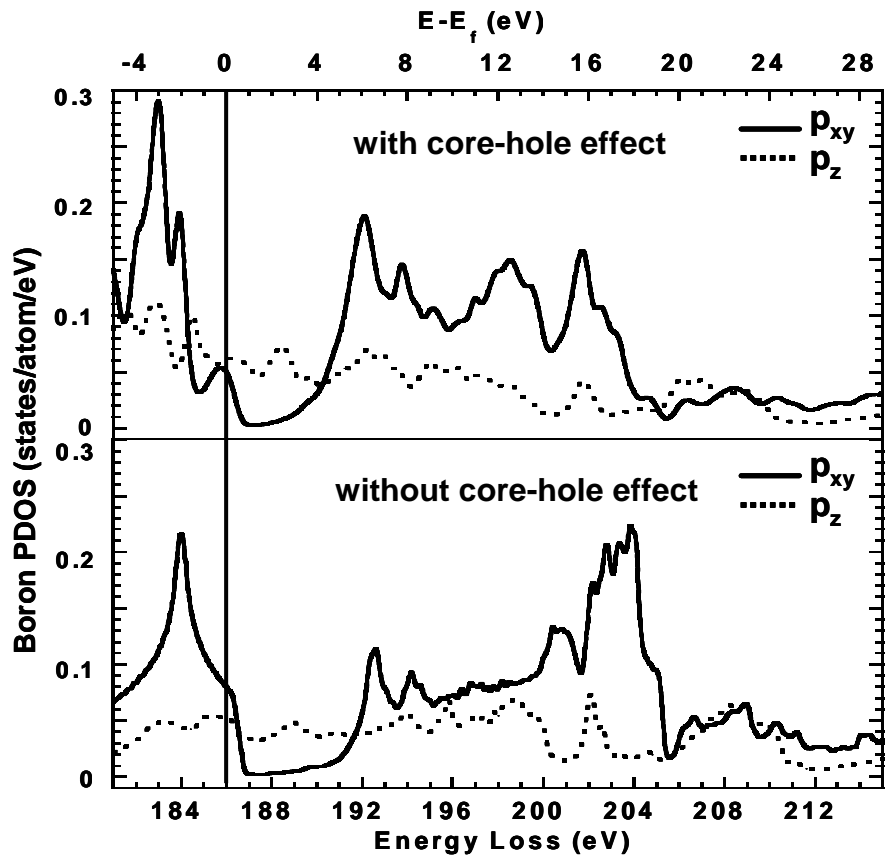

Figure 2:

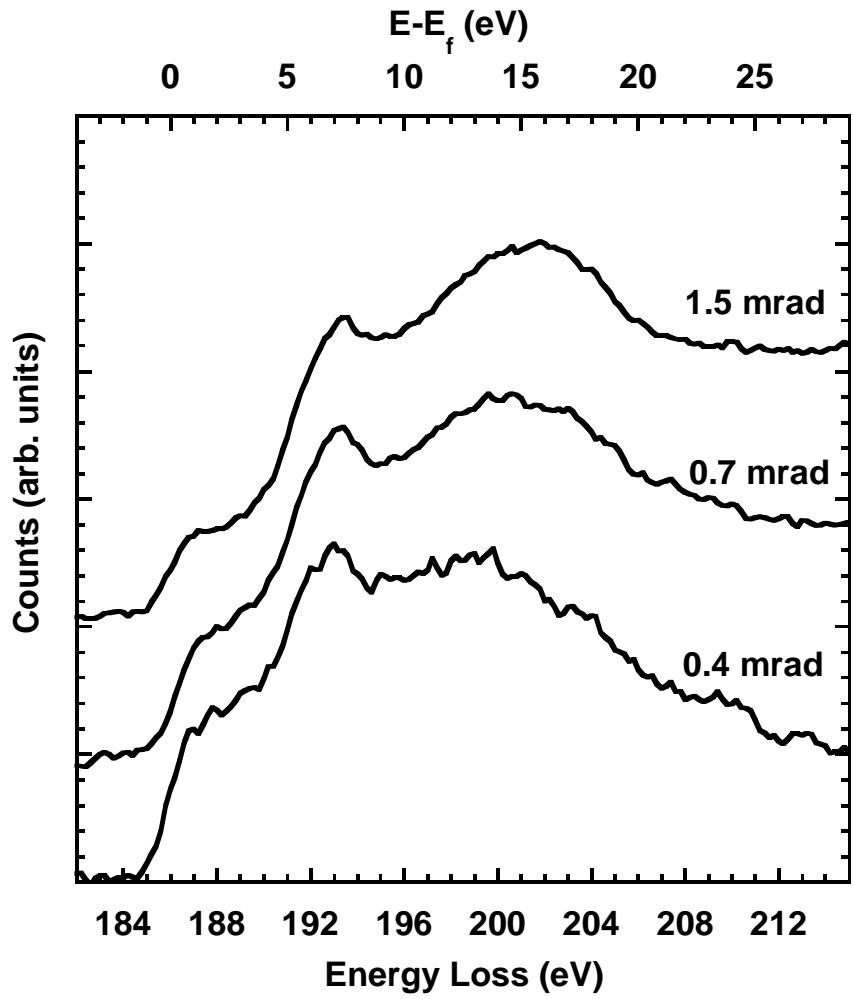

Figure 3:

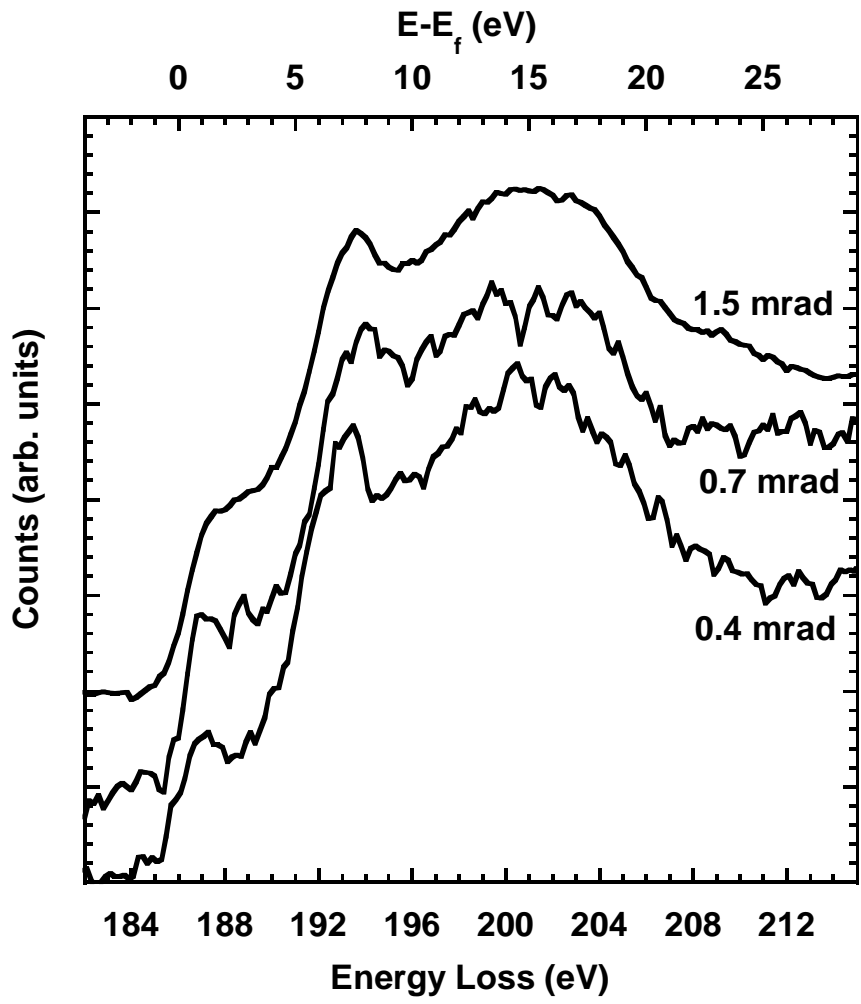

Figure 4:

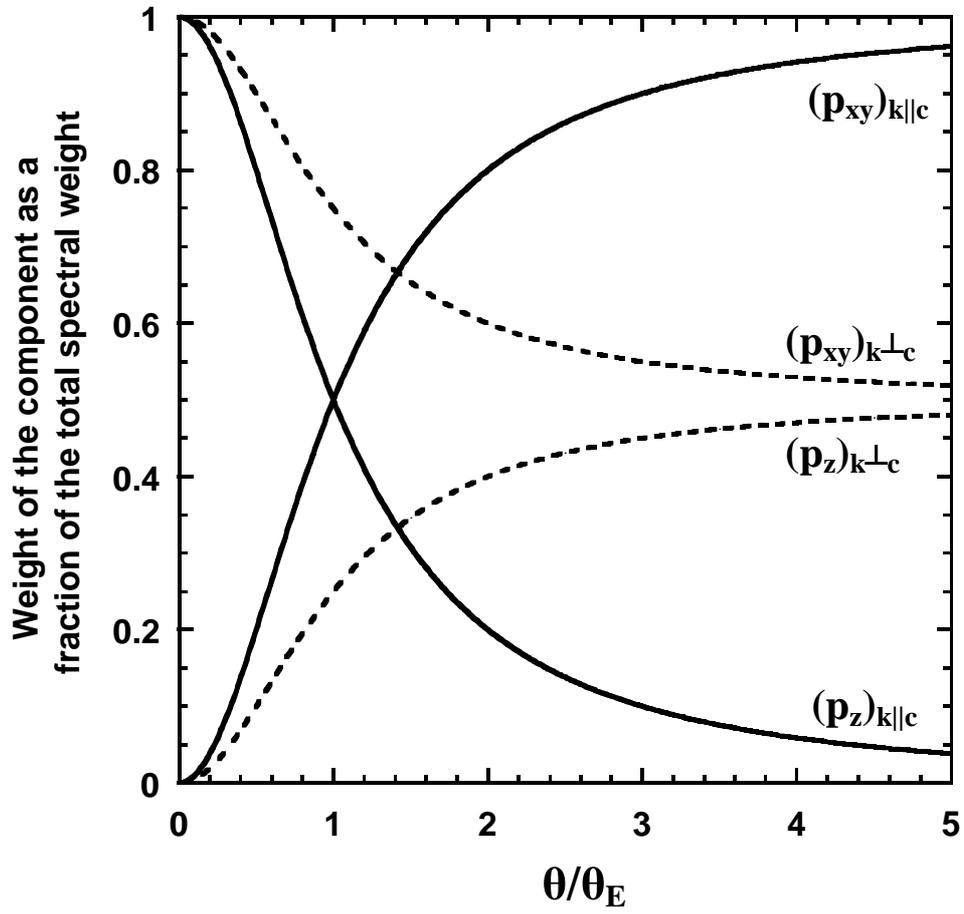

Figure 5:

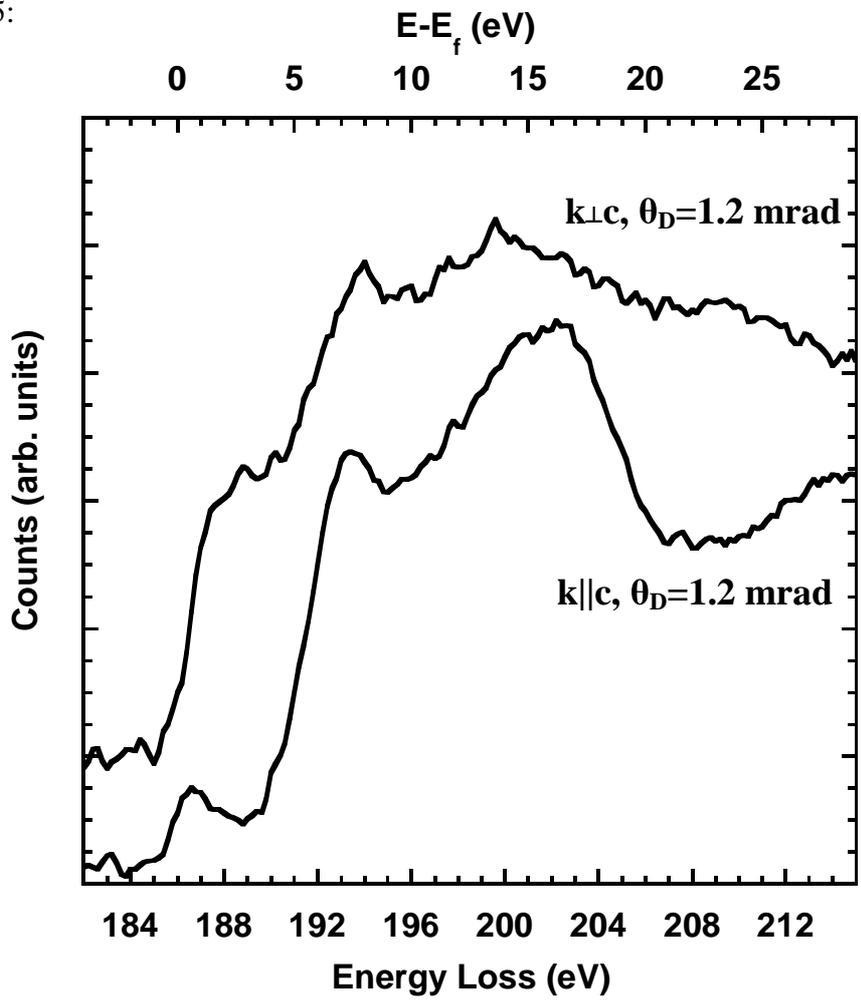